\documentclass[twocolumn,aps,prl,superscriptaddress]{revtex4-2}

\usepackage{amsmath}
\usepackage{graphicx}
\usepackage{MnSymbol}
\usepackage{parskip}
\usepackage{makecell}
\usepackage{booktabs}
\usepackage{multirow}

\begin{document}

\title{Tire tread block dynamics}

\author{N. Miyashita}
\affiliation{The Yokohama Rubber Company, 2-1 Oiwake, Hiratsuka, Kanagawa 254-8601, Japan}

\author{B.N.J. Persson}
\affiliation{State Key Laboratory of Solid Lubrication, Lanzhou Institute of Chemical Physics, Chinese Academy of Sciences, 730000 Lanzhou, China}
\affiliation{Peter Gr\"unberg Institute (PGI-1), Forschungszentrum J\"ulich, 52425, J\"ulich, Germany}
\affiliation{MultiscaleConsulting, Wolfshovener str. 2, 52428 J\"ulich, Germany}

\begin{abstract}
Temperature has a crucial influence on rubber friction and tire dynamics.
The temperature field in a rubber tread block is the sum of the {\it background temperature}
$T_0({\bf x},t)$, which varies slowly in time and space, and the {\it flash temperature} 
$\Delta T({\bf x},t)$, which in nonzero only close to the 
macroasperity contact regions, and which varies rapidly in time often on the millisecond time scale.
Here we study the motion of a single tire tread block and how it is influenced
by the flash temperature. We also present a theory and experimental results for 
the size of the macroasperity contact regions. 
In particular, we show that for a large enough nominal contact area,
in most cases the diameter $D$ of the macroasperity contact regions are nearly independent
of the elastic modulus and the nominal contact pressure.
\end{abstract}

\maketitle

\setcounter{page}{1}
\pagenumbering{arabic}




\begin{figure}
\includegraphics[width=0.27\textwidth,angle=0.0]{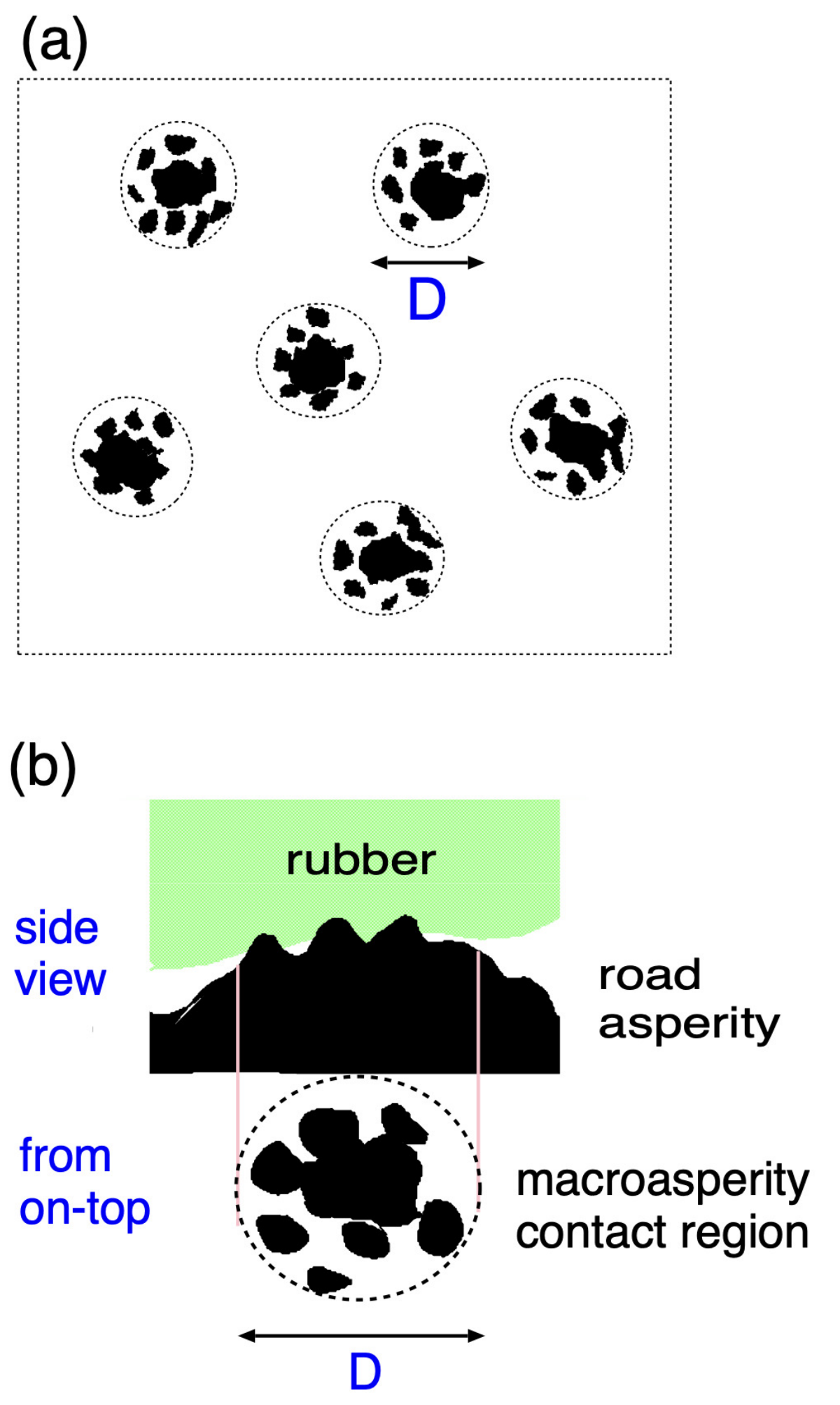}
\caption{\label{MacroAsperityTopSide.pdf}
If the nominal contact pressure is small enough
the contact between a rubber block and a road surface 
consist of well-separated macroasprity contact regions. Each macroasperity contact region
consist of closely spaced microasperity regions. When calculating the flash
temperature the frictional energy produced in a macroasperity contact 
region is smeared out within the macroasperity contact area to form a laterally uniform heat source.
This is a good approximation unless the sliding speed is very high
where the heat diffusion is to slow too smear out the temperature profile laterally.
}
\end{figure}

{\bf 1 Introduction}

Rubber friction is important in very many applications such as for the tire-road 
and shoe-road interaction, for the conveyor belt drive physics, 
and for the friction between the rubber stopper and the barrel in syringes. 
Most model studies of rubber friction consider rectangular blocks of rubber sliding on rough surfaces. 
The result of such studied have been compared to theories of rubber friction\cite{Gert,Gert1,Persson1}. 

When a rubber block slides on a rough surface most of the frictional energy is converted
into heat\cite{Persson1,Persson2,Persson3,Persson4}. Rubber friction depends sensitively on the temperature and the friction force
for a given sliding speed is maximal for a particular temperature. This is well known for racing tires
where the friction for sliding speeds of order 1 m/s is maximal for tire temperatures of order 100 C.
For this to be the case the rubber glass transition temperature must be $T_{\rm g} \approx - 10^\circ {\rm C}$,
while for rubber for normal passenger car tires typically $T_{\rm g} \approx -40^\circ {\rm C}$.

The temperature field in a tread block can be written as $T({\bf x},t) = T_0({\bf x},t) + \Delta T({\bf x},t)$.
The {\it background} temperature $T_0({\bf x},t)$ varies slowly in space and time while the {\it flash} temperature
$\Delta T({\bf x},t)$ varies very fast in space and time. $\Delta T({\bf x},t)$ in non-zero only close to the asperity contact regions
so very localized in space, see Fig. \ref{MacroAsperityTopSide.pdf}. 
If the {\it macroasperity} contact regions have the diameter $D$ then the flash temperature
occur on the time-scale $D/v \approx 1 \ {\rm ms}$ if $D\sim 1 \ {\rm mm}$ and the sliding speed $v \sim 1 \ {\rm m/s}$.
The background temperature is the sum of the cumulative flash temperature and a contribution from the 
time-dependent macroscopic viscoelastic deformation of the thread blocks and other parts of the tire. The background temperature depends on 
the driving history (cornering, acceleration, breaking) and on the external conditions (air and road temperature, humidity),
while the flash temperature depends on the instantaneous slip velocity, and only weakly on the external conditions.

In most cases the area of real contact is much smaller than the nominal contact area
and in these cases the friction force $F_{\rm f}$ is proportional to the normal force $F_{\rm N}$ and the
friction coefficient $\mu$, defined by $F_{\rm f} = \mu F_{\rm N}$, independent of
$F_{\rm N}$. The physical reason for this is that as the load increases the number of macroasperity contacts
increases proportional to the load while the average size of the macroasperity contact regions stays (nearly) constant.

For tires it is often found that $F_{\rm f}/F_{\rm N}$ depend on $F_{\rm N}$
but this is usually an indirect effect: When $F_{\rm N}$ increases the rubber background temperature increases,
which modify $F_{\rm f}$. Also, the length of the tire footprint in the sliding direction increases with increasing
$F_{\rm N}$, which also affect the friction force.

Rubber friction at time $t$ depends on the history $t' <t$ of the sliding motion because of the 
(slow) variation in the background temperature.
The flash temperature depends also on the sliding history but only during the (usually) very short
time it takes for the rubber to slide a distance of order the diameter $D$ of the macroasperity contact
regions. The aim of this study is to illustrate this effect with theory results for a realistic situation.
We also present theory and experimental results for the size of the macroasperity contact regions.

\vskip 0.3cm
{\bf 2 Tread block dynamics}

\begin{figure}
\includegraphics[width=0.47\textwidth,angle=0.0]{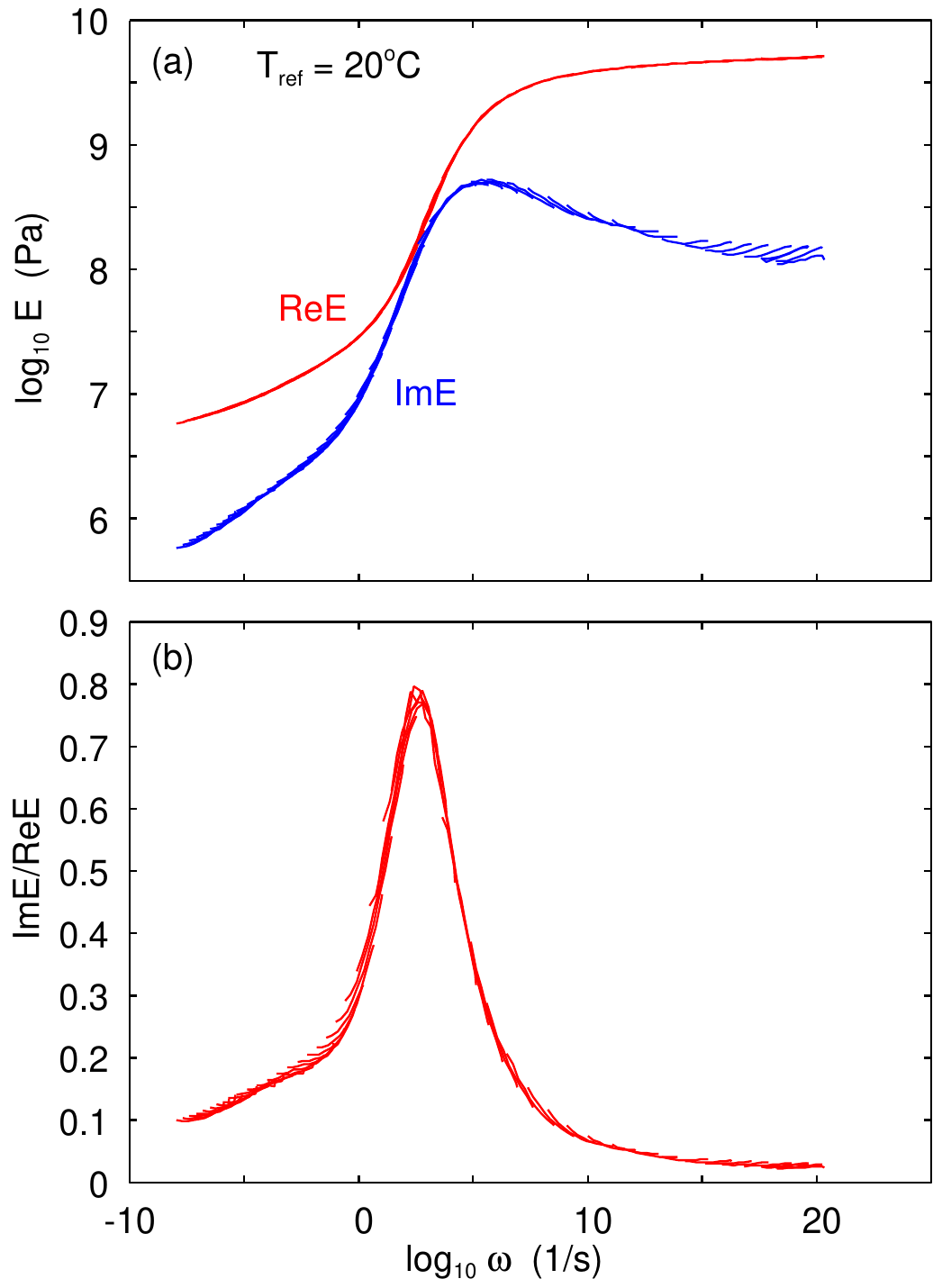}
\caption{\label{Emodulus.pdf}
(a) The real and imaginary part of the viscoelastic modulus $E(\omega)$ as a function of frequency
(log-log scale) and (b) the ratio ${\rm Im}E/{\rm Re}E$ as a function of the logarithm of the frequency.
For a racing tire tread compound with the glass transition temperature $T_{\rm g} = - 15^\circ {\rm C}$.
}
\end{figure}

We consider a tread rubber block of a racing compound sliding on a concrete surface. The small-strain (linear response)
viscoelastic modulus of the rubber is shown in Fig. \ref{Emodulus.pdf}. The rubber has the glass transition temperature 
$T_{\rm g} \approx - 15^\circ {\rm C}$ which is typical for racing compounds. As road surface we use the same concrete surface
as used in many earlier studies. The surface roughness power spectrum of the concrete surface was reported on
in Ref. \cite{Persson1,Cq} and in the Sec. 3.   

\begin{figure}
\includegraphics[width=0.47\textwidth,angle=0.0]{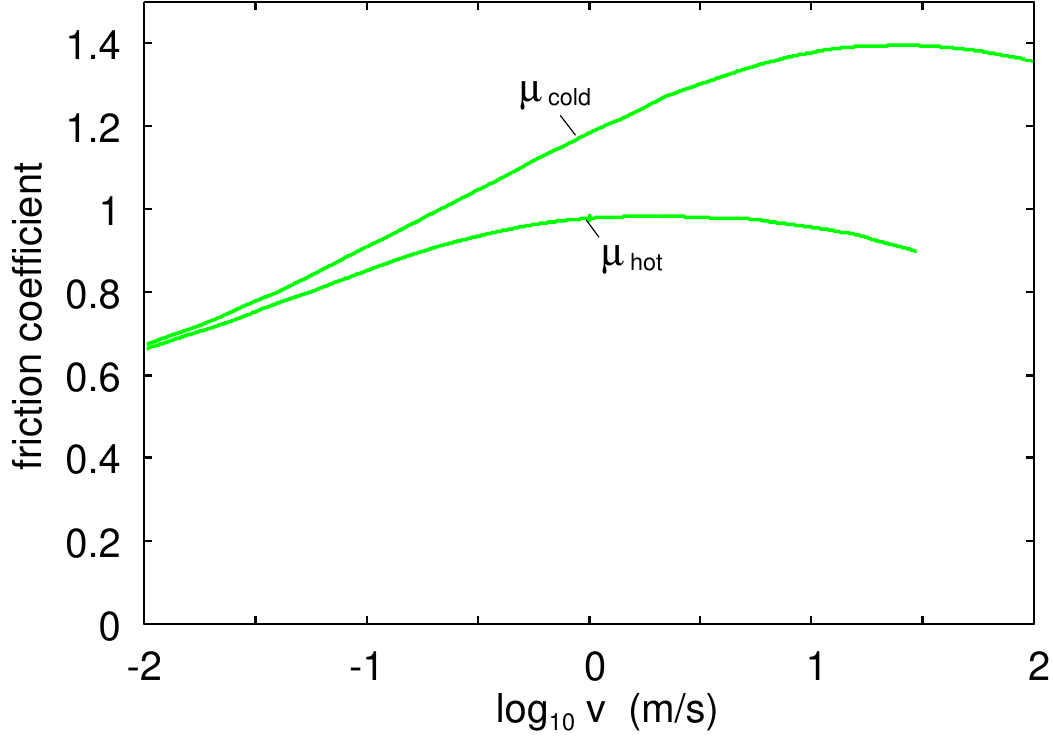}
\caption{\label{1logv.2mu.racing.pdf}
The calculated viscoelastic contribution to the friction coefficient as a function of the logarithm of the 
sliding speed for the temperature $T=60^\circ{\rm C}$. 
For the racing compound sliding on a concrete surface. The upper curve denoted $\mu_{\rm cold}$
is without the flash temperature and the lower curve denoted $\mu_{\rm hot}$ is with the flash temperature.
}
\end{figure}

There are two contributions to the friction force for rubber sliding on clean 
surfaces, namely a viscoelastic contribution $\mu_{\rm visc} F_{\rm N}$ from the 
deformation of the rubber in the asperity contact regions,
and a contribution $\mu_{\rm ad} F_{\rm N}$ from shearing the area of real contact\cite{Persson2,Schall,Persson5}. For tires on clean
road surfaces both contributions are roughly equally important but they depend on the sliding speed in different ways. 
The contribution from the area of contact can be written as 
the product between the area of real contact $A$ and a frictional shear stress $\tau$.
We refer to $\mu_{\rm ad}$ as the adhesive contribution as for clean surfaces it is assumed to result
from molecular stick-slip processes where segments of the rubber chain molecules bind to the countersurface,
followed by stretching of the chain and bond breaking, 
where the stored up elastic stretching energy is converted into heat\cite{Persson5}.

The Persson contact mechanics and rubber friction theory predict $\mu_{\rm visc}$ and the
area of real contact $A$ but the dependency of $\tau (v,T)$ on the velocity
and the temperature $T$ must be obtained from other theories\cite{Schall,Persson5} or from experiments.
Here we will only consider the viscoelastic contribution to the friction, which is most important 
on contaminated road surfaces,
but the basic picture is valid also when including the adhesive contribution.

\begin{figure}
\includegraphics[width=0.47\textwidth,angle=0.0]{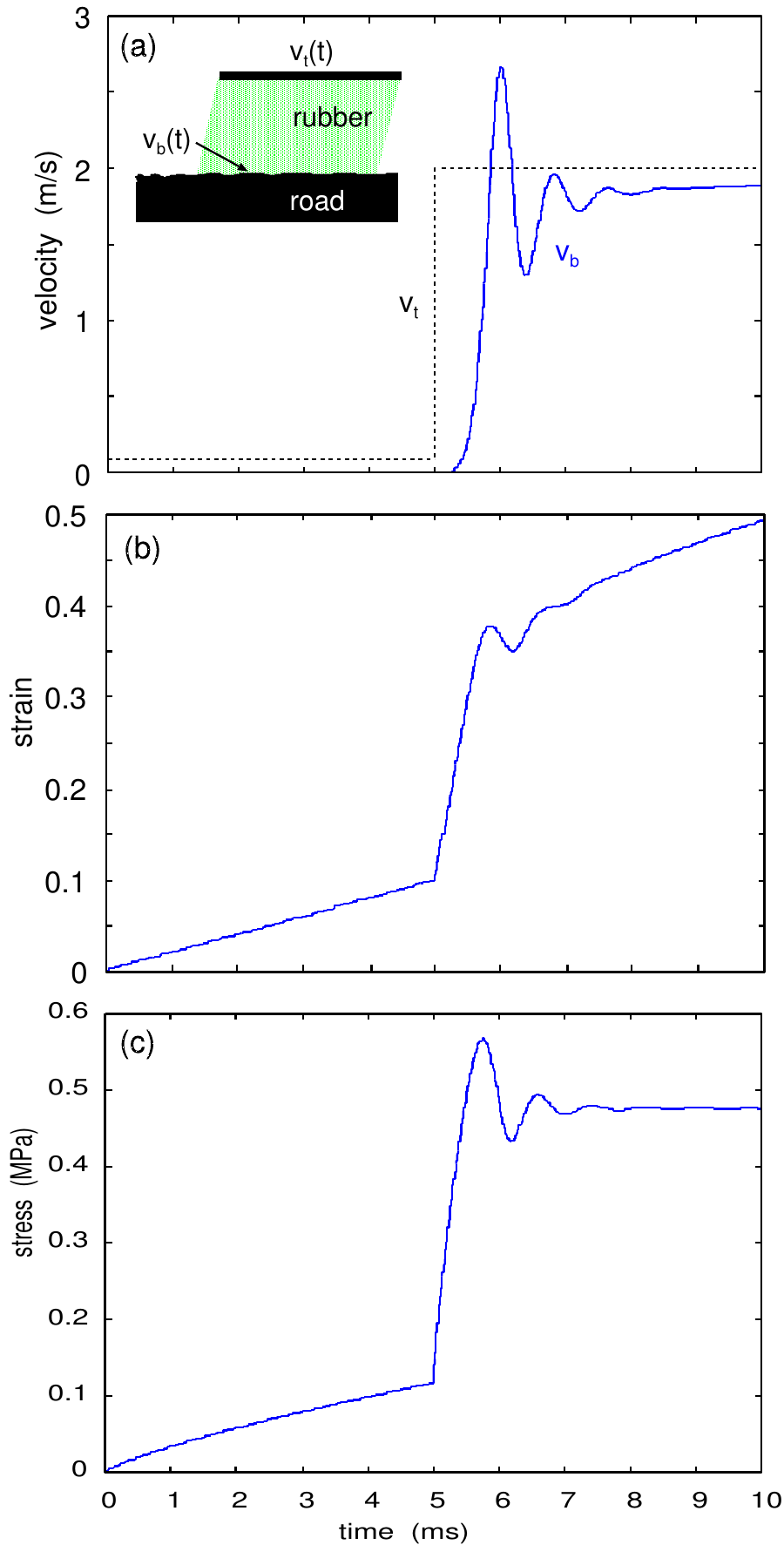}
\caption{\label{one.pdf}
Calculated results for a rubber tread block sliding on a concrete surface.
The upper surface of the rubber block is glued to a rigid surface which moves with the speed
$v_{\rm t}=0.1 \ {\rm m/s}$ for the time period $0 < t < 5 \ {\rm ms}$ and with $v_{\rm t}=2 \ {\rm m/s}$ for the time period $5 < t < 10 \ {\rm ms}$.
(a) shows the time dependency of the top surface velocity $v_{\rm t}$ (dashed line) and the 
velocity $v_{\rm b}$ of the bottom surface, which is in contact with the concrete surface. 
(b) shows the stain and (c) the shear stress in the rubber block as a function of time.
}
\end{figure}

\begin{figure}
\includegraphics[width=0.47\textwidth,angle=0.0]{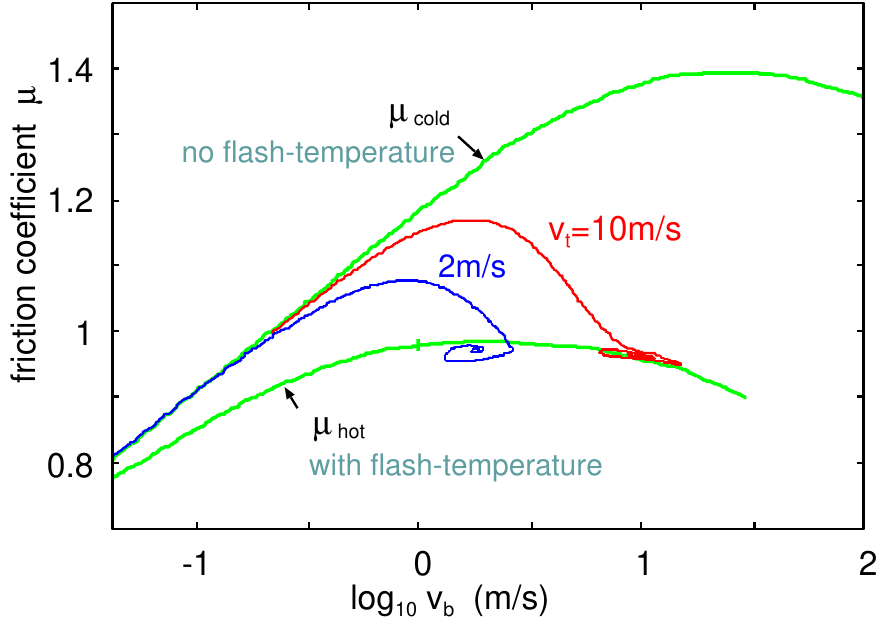}
\caption{\label{two.pdf}
The dependency of the effective friction coefficient on the velocity $v_{\rm b}$ of the bottom surface
of the rubber block. The green lines shows the steady-state friction coefficients without (upper curve) and with (lower curve)
the flash temperature. The blue curve is the friction coefficient experienced by the rubber block
for the sliding case in Fig. \ref{one.pdf}, and the red curve for a similar case but where
the sliding speed increases from $0.1$ to $10 \ {\rm m/s}$ at $t=5 \ {\rm ms}$ instead of $2 \ {\rm m/s}$ as in Fig. \ref{one.pdf}.
The dynamic friction coefficient first follows the cold-branch $\mu_{\rm cold}$ but after sliding a distance of order
the size of the macroasperity contact, which is needed to fully develop the flash temperature, 
it transition to the hot branch $\mu_{\rm hot}$.
}
\end{figure}

Fig. \ref{1logv.2mu.racing.pdf}
shows the calculated viscoelastic contribution to the friction coefficient as a function of the logarithm of the 
sliding speed  for the racing compound sliding on a concrete surface. 
The background temperature $T_0=60^\circ{\rm C}$. 
The results are obtained using the Persson contact mechanics and rubber friction theory and the basic
equations used are presented in Ref. \cite{Persson2,Persson3}. In the theory use an effective large-strain viscoelastic modulus
(strain $\sim 0.5$) as both the macroscopic shear deformations of the rubber block and the strain in the rubber road asperity
contact regions is of order unity while the mastercurve shown in Fig.
\ref{??} was obtained for very low strain ($\approx 4\times 10^{-4}$) where the linear-response assumption is accurate. 
The upper curve in the figure denoted $\mu_{\rm cold}$ 
is without the flash temperature, and the lower curve denoted $\mu_{\rm hot}$ is with the flash temperature.

The results in Fig. \ref{1logv.2mu.racing.pdf} are for stationary sliding. However, when a tread block moves through the tire-road
footprint the sliding speed is highly non-uniform. For small tire slip there is no slip between the rubber and the road
surface until the shear stress becomes big enough to overcome the effective static frictional stress. To simulate this we
consider a tread block squeezed against the road surface with the nominal pressure $0.5 \ {\rm MPa}$. The upper surface
of the tread block first moves with the speed $v_{\rm t} = 0.1 \ {\rm m/s}$ for $5 \ {\rm ms}$ after which the speed is increased to
$v_{\rm t} = 2 \ {\rm m/s}$. In tire applications only the second phase ($t>5 \ {\rm ms}$) is relevant but we consider
here a more general case as it illustrate the physics better. For the system considered here (racing compound on the concrete surface)
the diameter of the macroasperity contact regions is $D \approx 1 \ {\rm mm}$ as observed using pressure sensitive paper,
and also predicted by the theory (see Sec. 3). We note that 
if the relative contact area $A/A_0 \lesssim 0.1$ (which is obeyed in most tire applications) 
then the size of the macroasperity contact regions are nearly unchanged as the sliding speed increases, even though the magnitude of the 
effective viscoelastic modulus increases with sliding speed. 
Furthermore, the size of the macroasperity contact regions is also neraly independent of the nominal contact pressure. 
However, these counterintuitive results are valid only if the nominal rubber-road contact area $A_0$ is large enough, see Sec. 3.

The shear deformation $[x_{\rm b}(t)-x_{\rm t}(t)]/d$, where $x_{\rm t}$ and $x_{\rm b}$ are the $x$-coordinate of the
top and bottom surface of the tread block, is related to the shear stress $\tau (t) = \sigma_{xz}(t)$ using\cite{Persson3} 
$$\tau (t) = \int_0^t dt' \, K(t-t') [x_{\rm b} (t')-x_{\rm t} (t')]. $$
Here $K(t) = G(t)/d$, where $d$ is the thickness of the tread block, and
$$G(t) = {1 \over 2 \pi} \int_{-\infty}^\infty d\omega \, G(\omega) e^{-i \omega t}, $$
where 
$$G(\omega) = {E(\omega)\over 2[1+\nu (\omega)]}$$ 
is the shear modulus. Note that the memory-spring $K(t)$ describe both elasticity and damping as contained in the
shear modulus $G(\omega)$ (see Ref. \cite{Persson3} for more details).

Fig. \ref{one.pdf} shows the calculated results.
In (a) we shows the time dependency of the top surface velocity $v_{\rm t}$ (dashed line) and the 
velocity $v_{\rm b}$ of the bottom surface of the rubber block, which is in contact with the concrete surface. 
Fig. \ref{one.pdf}(b) shows the stain and (c) the shear stress in the rubber block as a function of time.

Fig. \ref{one.pdf}(a) shows that the rubber block does not slip on the concrete surface during the first
$5 \ {\rm ms}$ where the speed of the top surface is $v_{\rm t} = 0.1 \ {\rm m/s}$. During this time the stain
increases linearly with time $\epsilon = v_{\rm t} t/d$ where the tread block height $d = 0.5 \ {\rm cm}$.
For $t=5 \ {\rm ms}$ the strain $\epsilon = 0.1 \times 0.005 /0.005 = 0.1$ as shown in \ref{one.pdf}(b).
During this time period, as shown in Fig. \ref{one.pdf}(c), the shear stress depends slightly non-linear on 
time due to viscoelastic relaxation. After the switch to the sliding speed $v_{\rm t} = 2 \ {\rm m/s}$ at time
$t=5 \ {\rm ms}$ the strain and stress increases fast and slip at the rubber-concrete interface start at
$t \approx 5.3 \ {\rm ms}$. The velocity of the bottom of the rubber block perform damped oscillations
for $t > 6 \ {\rm ms}$ which also shows up as oscillations in the strain and the shear stress. The strain continue to increase up to
the largest time studied, as also manifested in the slip velocity $v_{\rm b}$ which is less than the driving velocity $v_{\rm t}$
even at $t=10 \ {\rm ms}$. This is again related to viscoelastic relaxation in the tread block.

\begin{figure}
\includegraphics[width=0.47\textwidth,angle=0.0]{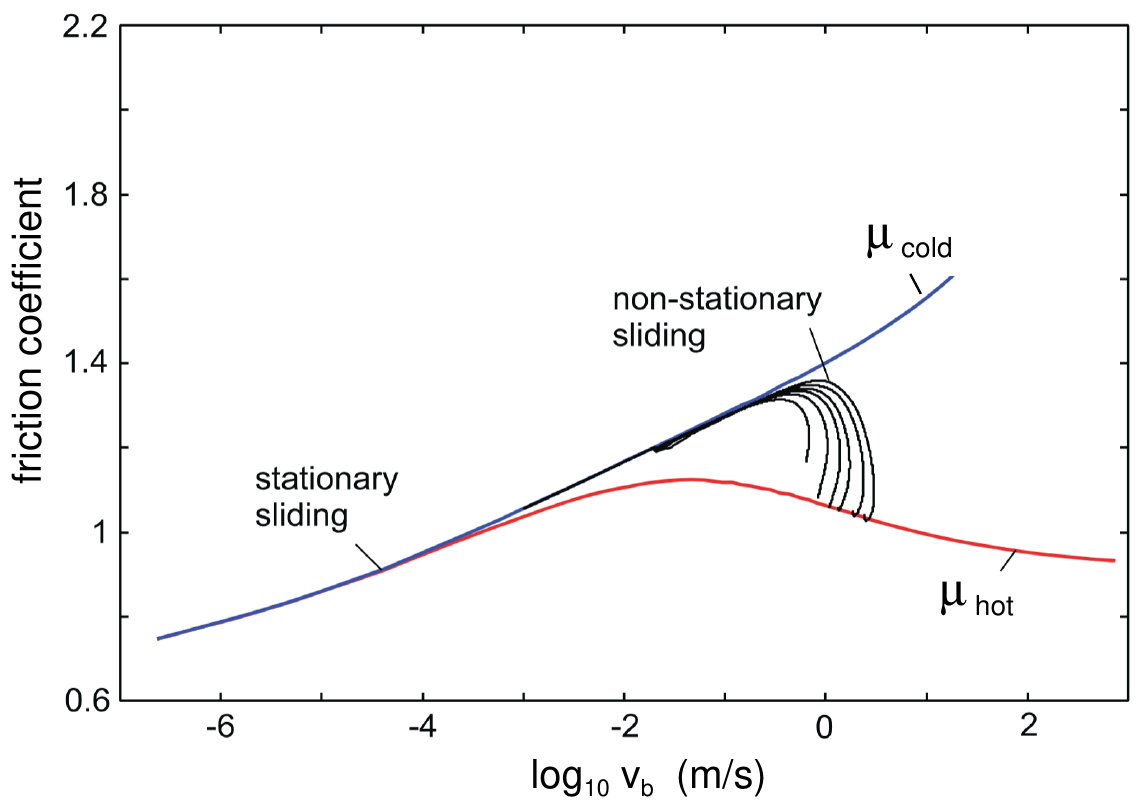}
\caption{\label{TireSlipMany.pdf}
Red and blue lines: the kinetic friction coefficient
(stationary sliding) as a function of the logarithm (with 10 as basis)
of the slip velocity $v_{\rm b}$. The blue line denoted $\mu_{\rm cold}$ is without the
flash temperature while the red line denoted $\mu_{\rm hot}$ is with the flash
temperature. Black curves: the effective friction coefficient experienced by a
tread block as it goes through the footprint. For a car velocity
$27 \ {\rm m/s}$ and for several slip values $s=0.005 $, 0.0075, 0.01, 0.03, 0.05,
0.07 and 0.09. Note that the friction experienced by the tread block
first follows the $\mu_{cold} (v)$ rubber branch of the steady state kinetic
friction coefficient and then, when the block has slip a distance of the
order of the diameter of the macroasperity contact region, it follows
the $\mu_{\rm hot}(v)$ rubber friction branch.
}
\end{figure}

The most important result of this study is shown in Fig. \ref{two.pdf}. This figure
shows the dependency of the effective friction coefficient on the velocity $v_{\rm b}$ of the bottom surface
of the rubber block. The green lines shows the steady-state friction coefficients without (upper curve) and with (lower curve)
the flash temperature. The blue curve is the friction coefficient experienced by the rubber block
for the sliding case in Fig. \ref{one.pdf}, and the red curve for a similar case but where
the sliding speed increases from $0.1$ to $10 \ {\rm m/s}$ at $t=5 \ {\rm ms}$ instead of $2 \ {\rm m/s}$ as in Fig. \ref{one.pdf}.
The dynamic friction coefficient first follows the cold-branch $\mu_{\rm cold}$ but after sliding a distance of order
the size $D$ of the macroasperity contact, which is needed to fully develop the flash temperature, 
it transition to the hot branch $\mu_{\rm hot}$. Similar dynamic friction coefficients is found in tire dynamics simulations as
shown in Fig. \ref{TireSlipMany.pdf}.

Fig. \ref{TireSlipMany.pdf} shows the dependency of the effective friction coefficient on the slip velocity $v_{\rm b}$ 
for a tire tread block as it pass though the tire-road footprint. 
The results are obtained using the tire model developed in Ref. \cite{Persson6,Persson7}.   
The blue and red curves in the figure show the $v_{\rm b}$-dependency of the
kinetic friction coefficient, for stationary sliding without 
and with the flash temperature, respectively.
The black curves show the effective friction during non-stationary 
sliding experienced by a rubber tread block during
braking at various tire slip $s=(v_{\rm R}-v_x)/v_x$ ($v_{\rm R}$ is the tire rolling velocity 
and $v_x$ the tire center of mass velocity). Note
that, because of the finite sliding distance which is necessary in order
to fully develop the flash temperature, the friction acting on
the tread block initially follows the blue curve corresponding
to cold rubber (i.e. negligible flash temperature). Thus, it is
not possible to accurately describe rubber friction with just
a static and a kinetic friction coefficient (as is often done
even in advanced tire dynamics computer simulation codes) or
even with a function $\mu (v_{\rm b})$ which depends on the instantaneous
sliding velocity $v_{\rm b}(t)$. 
Instead, the friction depends on $v_{\rm b}(t)$ for earlier times $t'<t$.
This memory effect is due to the dependency of the temperature on the
sliding history.

For the full tire model, Fig. \ref{TireSlipMany.pdf}
shows that there are no oscillations of the tire tread blocks in the tire-road
footprint. This may be due to the viscoelastic coupling between the tread blocks.

We have found that it is possible to describe the dependency of the friction coefficient on time
during non-stationary sliding using a simple friction law.
If $x_{\rm b}(t)$ denote the slip distance between the rubber block and a road surface then\cite{Persson6,Persson7}
$$\mu (t) = \mu_{\rm cold}(v_{\rm b}(t),T_0) e^{-\alpha x_{\rm b}(t)/D}$$
$$ + \mu_{\rm hot}(v_{\rm b}(t),T_0)\left [1-e^{-\alpha x_{\rm b}(t)/D}\right ],$$
where $\alpha \approx 5$, gives tread block motion in good agreement with the full numerical calculation.
Using this friction law in tire simulations gives nearly the same result as using the full
dynamic friction description.

\begin{figure}
\includegraphics[width=0.47\textwidth,angle=0.0]{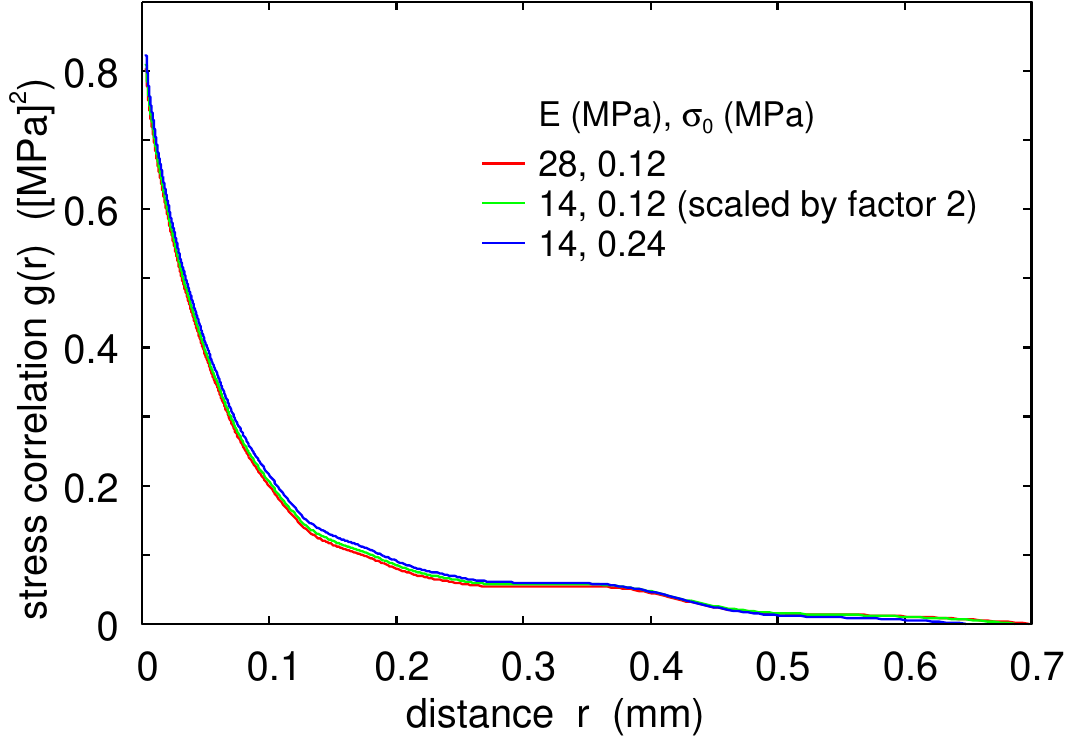}
\caption{\label{1r.2StressCorrelation.pdf}
The stress correlation function $g(r)$ as a function of the distance
$r=|{\bf x}-{\bf x}'|$ for rubber block squeezed against the concrete surface. 
The red, green and blue curves are calculated
with the Young's modulus $E$ (in MPa) and the applied nominal contact pressure 
$\sigma_0$ (in MPa) given by $(E,\sigma_0) = (28,0.12)$ (red curve), $(14, 0.12)$ (green)
and $(14,0.24)$ (blue). The $g(r)$ for the green curve is scaled by a factor of
$2$. 
}
\end{figure}

\begin{figure}
\includegraphics[width=0.4\textwidth,angle=0.0]{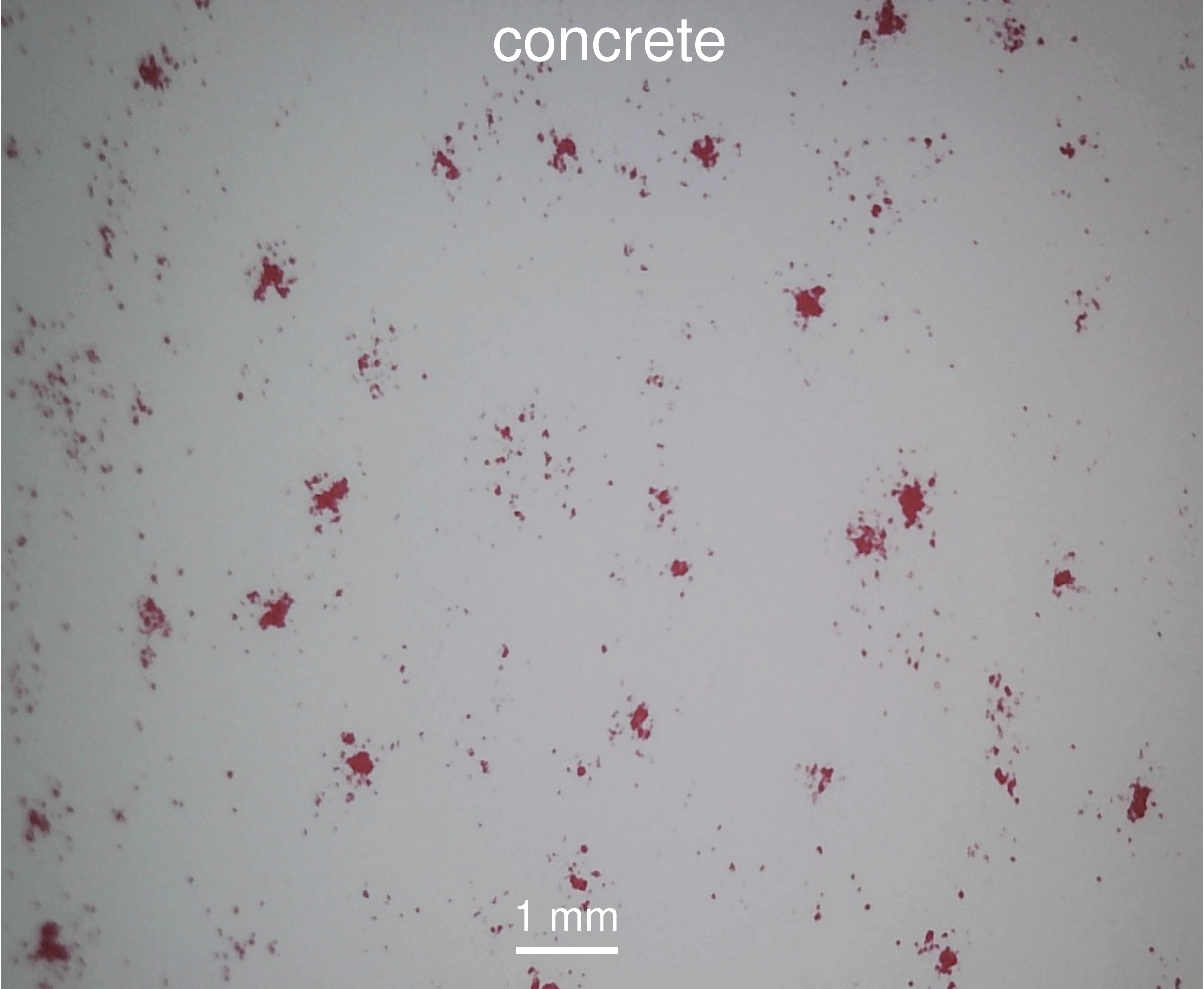}
\caption{\label{FIRST.pdf}
Picture of the contact regions between a rubber block and the concrete surface.
The nominal contact pressure $\sigma_0 \approx 0.12 \ {\rm MPa}$.
The pictures are obtained with a pressure-sensitive film
(Fujifilm, Super Low Pressure, $0.5-2.5 \ {\rm MPa}$ pressure range).
}
\end{figure}

\begin{figure}
\includegraphics[width=0.4\textwidth,angle=0.0]{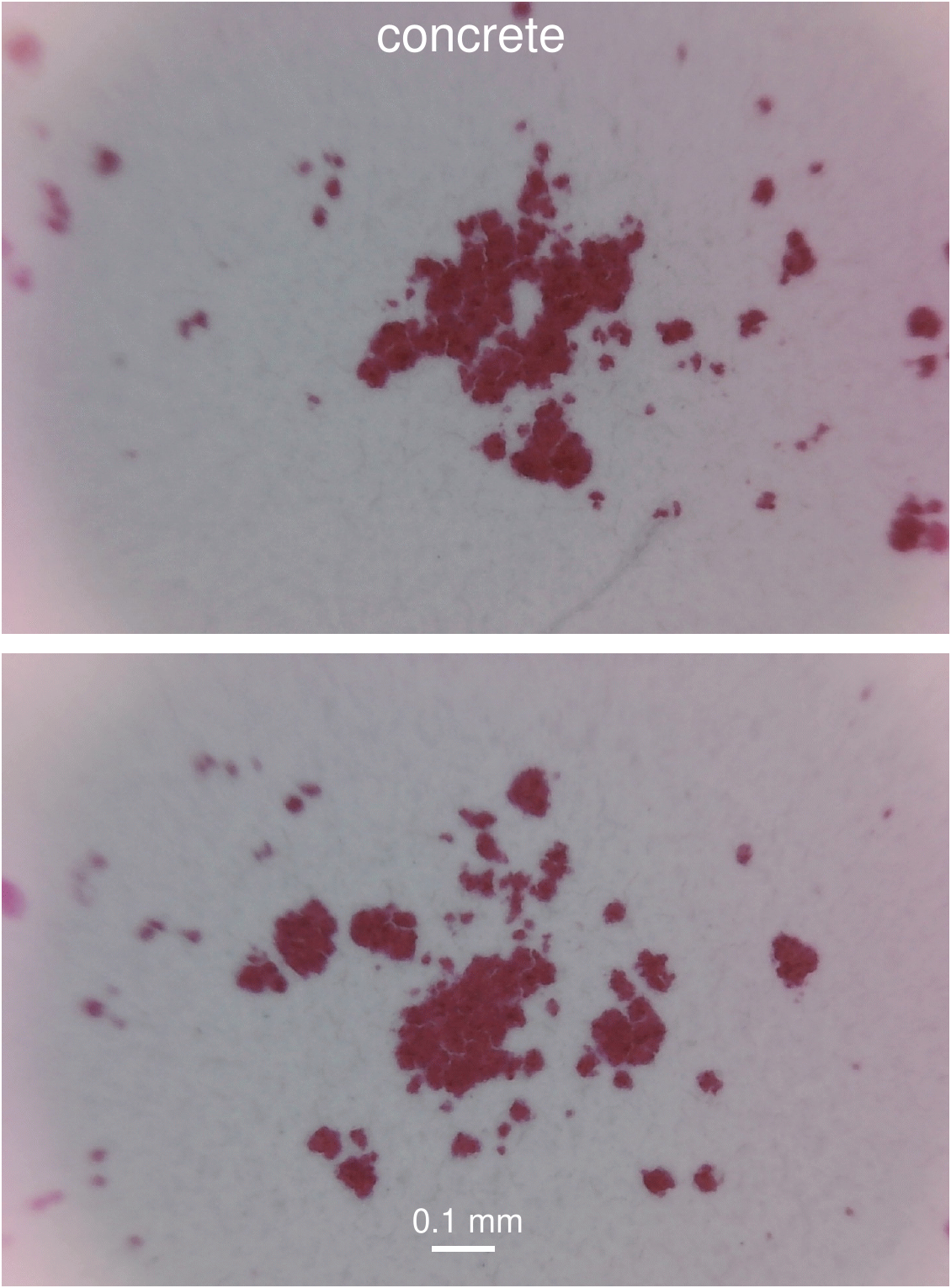}
\caption{\label{SECOND.pdf}
Magnified views of two contact regions from Fig. \ref{FIRST.pdf}.
}
\end{figure}

\begin{figure}
\includegraphics[width=0.4\textwidth,angle=0.0]{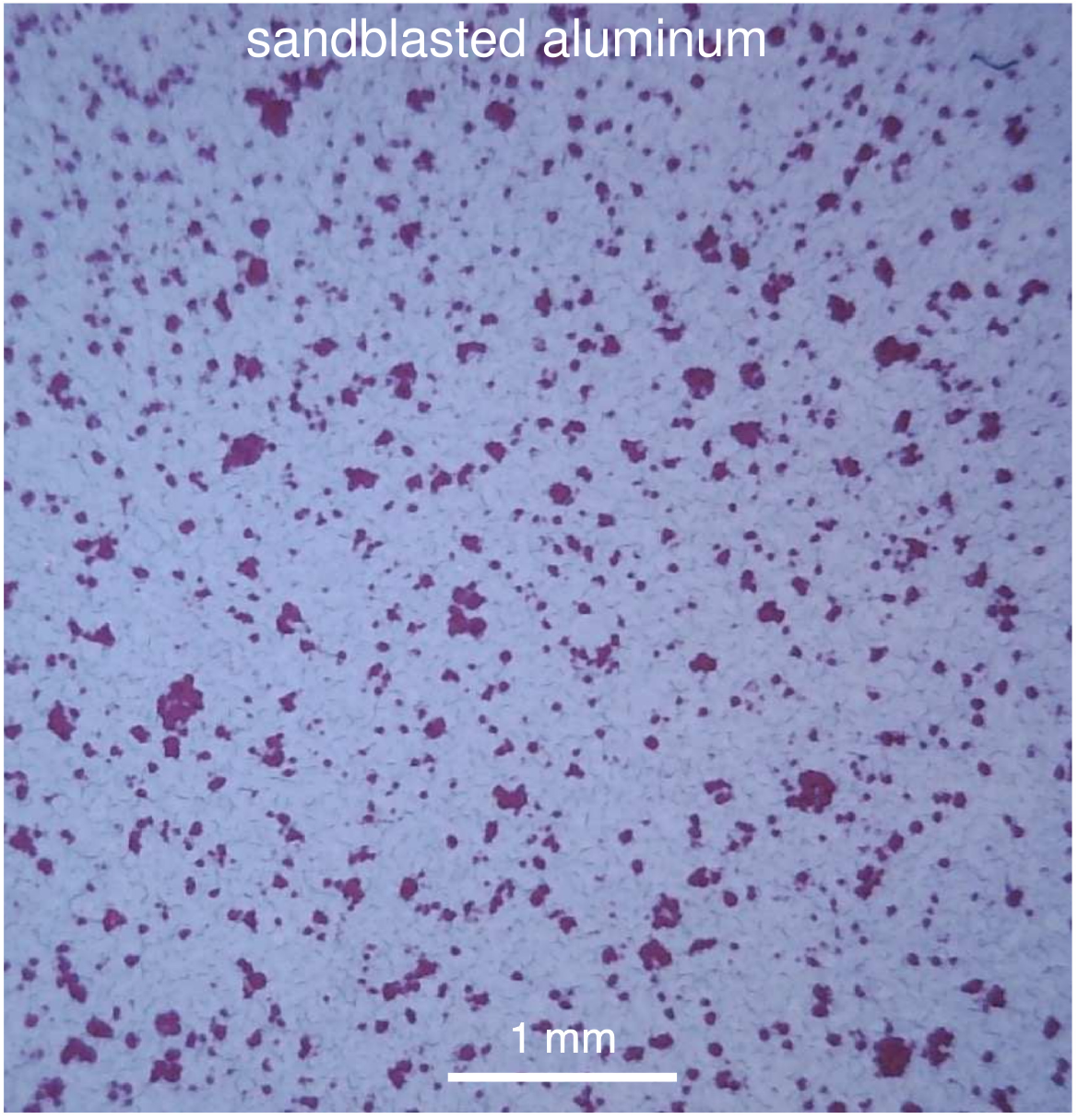}
\caption{\label{SandblastedAlBigWiev.pdf}
Picture of the contact regions between a rubber block and a sandblasted aluminum surface.
The nominal contact pressure is about $\sigma_0 \approx 1 \ {\rm MPa}$.
}
\end{figure}

\begin{figure}
\includegraphics[width=0.4\textwidth,angle=0.0]{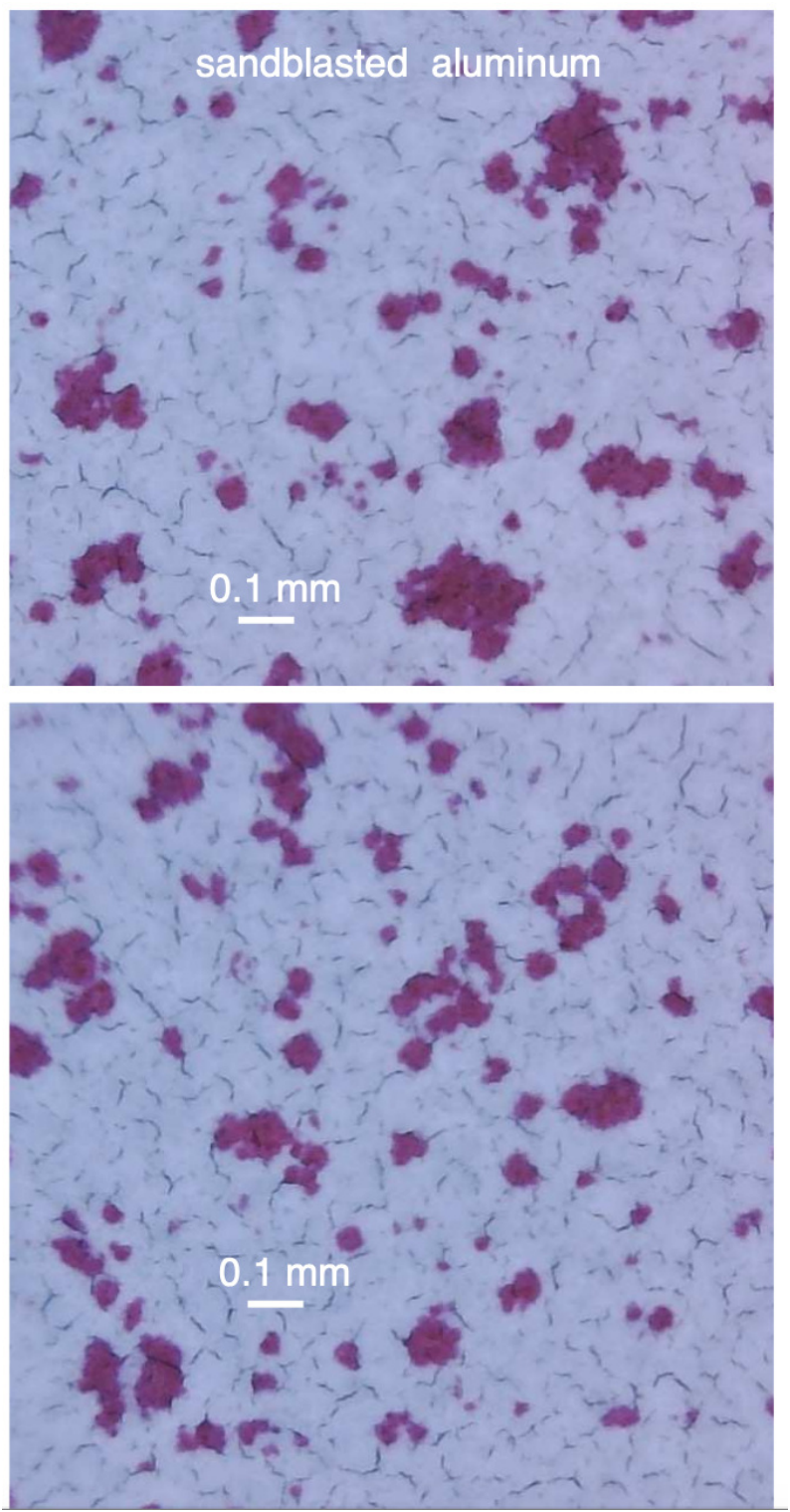}
\caption{\label{SandBlastedAl.png.pdf}
Magnified views of two contact regions for rubber block and the sandblasted aluminum surface.
}
\end{figure}

\begin{figure}
\includegraphics[width=0.4\textwidth,angle=0.0]{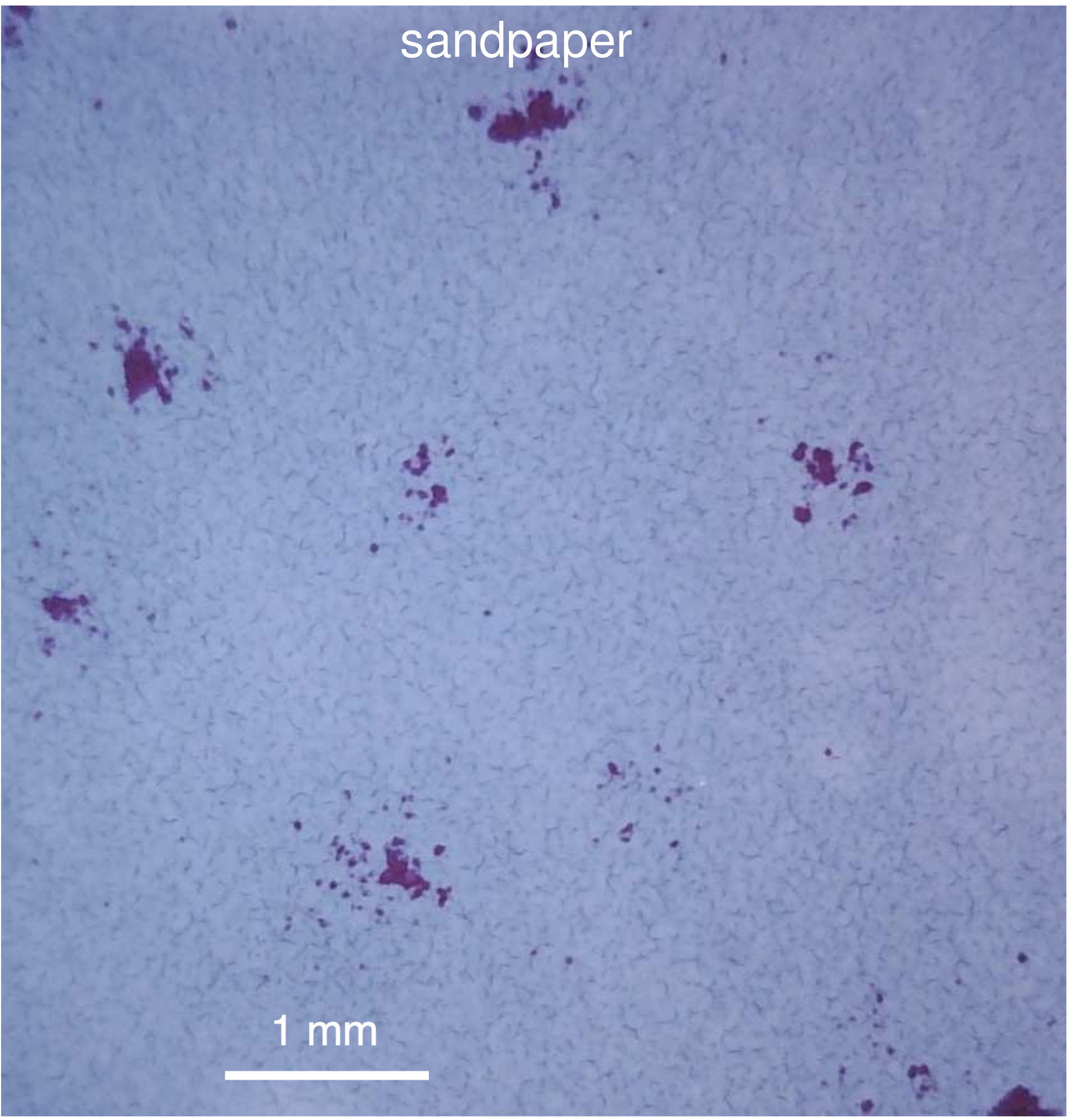}
\caption{\label{SandPaperBig.pdf}
Picture of the contact regions between a rubber block and a sandpaper P100 surface.
The nominal contact pressure is about $\sigma_0 \approx 0.2 \ {\rm MPa}$.
}
\end{figure}

\begin{figure}
\includegraphics[width=0.4\textwidth,angle=0.0]{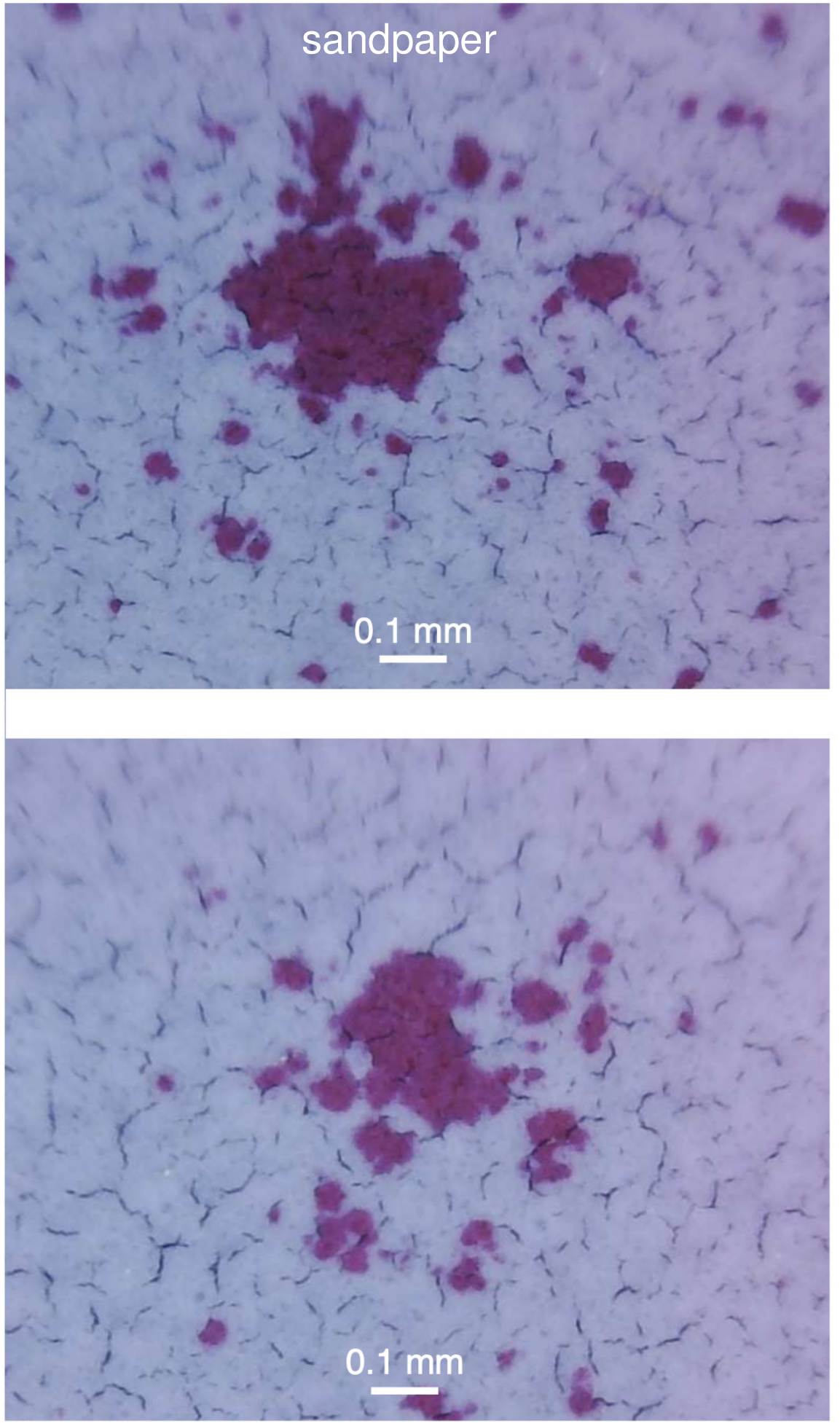}
\caption{\label{SandpaperPicMagnified.pdf}
Magnified views of two contact regions for rubber block and the sandpaper surface.
}
\end{figure}

\begin{figure}
\includegraphics[width=0.47\textwidth,angle=0.0]{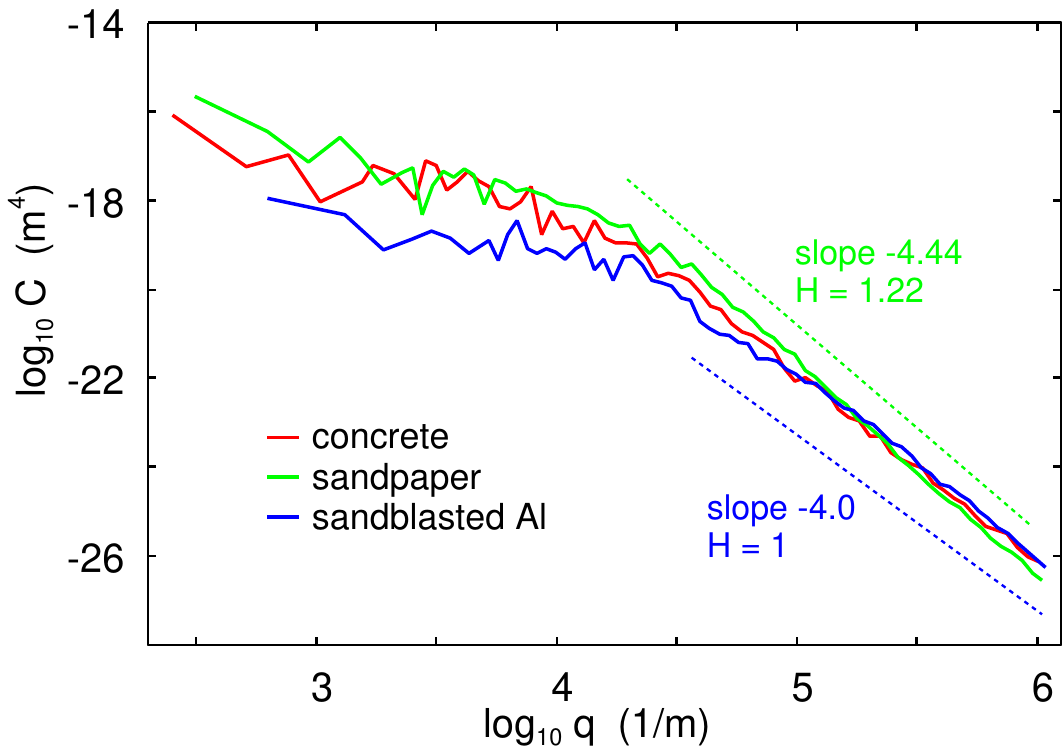}
\caption{\label{1logq.2logC.Concrete.Sandpaper.SandblastedAl.pdf}
Surface roughness power spectra of the concrete, sandpaper and sandblasted aluminum surface.
In the linear region the slope of the power spectra is $\approx -4.4$ for the
sandpaper surfaces and $-4.0$ for the sandblasted aluminum surface.
}
\end{figure}

\begin{figure}
\includegraphics[width=0.47\textwidth,angle=0.0]{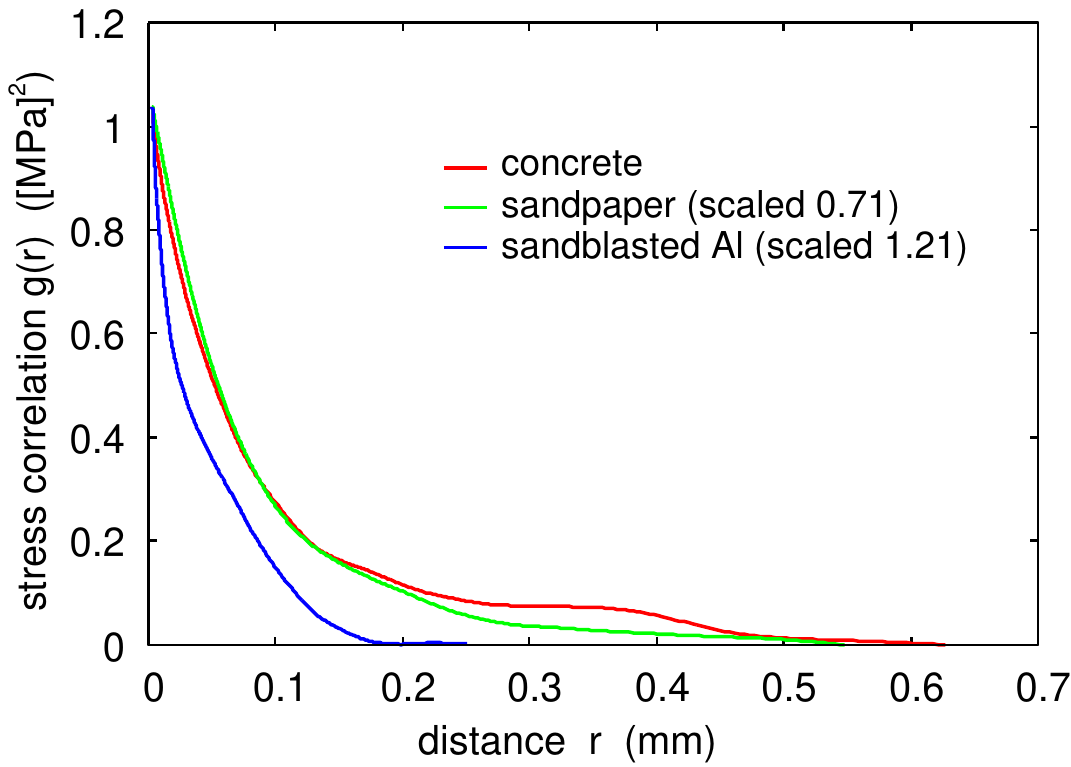}
\caption{\label{1radius.2StressCorrelation.Concrete.sandpaper.sandblastedAl.pdf}
The stress correlation function $g(r)$ as a function of the distance
$r=|{\bf x}-{\bf x}'|$ for an elastic block squeezed against the concrete (red line), the sandpaper (green) and the sandblasted
aluminum (blue) surfaces.
With the Young's modulus $E = 14 \ {\rm MPa}$ and the applied nominal contact pressure 
$\sigma_0 = 0.3 \ {\rm MPa}$.
The $g(r)$ for the green curve is scaled by $0.71$ and for the blue curve with $1.21$.
}
\end{figure}

\vskip 0.3cm
{\bf 3 On the size of macroasperity contact regions}

The effective size of the macroasperity contact regions can be determined from the stress-stress correlation function. 
The basic idea is that the contact stress is non-zero only in the contact area so the decay of the stress-stress correlation function
with the distance between two points will reflect the lateral size of the contact regions. 
The size of the macroasperity contact regions  is determined mainly by long wavelength roughness components but for the ``internal''
structure (microasperity contact regions) of the macroasperity contact
regions depends also the short wavelength part of the surface roughness power spectrum.
The treatment below is inspired by the studies of M\"user et. al. on the
size of contact regions and its relation to the stress-stress correlation function\cite{Mu1,Mu2}.

Consider the stress-stress correlation function
$$g({\bf x},{\bf x}') = 
\langle \sigma ({\bf x}) \sigma ({\bf x}')\rangle - \langle \sigma ({\bf x}) \rangle 
\langle \sigma ({\bf x}') \rangle.$$
The applied stress is defined as $\sigma_0 = \langle \sigma ({\bf x}) \rangle$, and from now on, we assume the applied stress is subtracted 
from $\sigma ({\bf x})$, so that $\langle \sigma ({\bf x}) \rangle = 0$.
For surfaces with isotropic roughness $g({\bf x},{\bf x}')$ depends only on $r=|{\bf x}-{\bf x}'|$ and we denote it by
$g(r)$. For this case we have shown in Ref. \cite{stresscorrelation}
$$\langle \sigma ({\bf q}) \sigma (-{\bf q})\rangle = {A_0\over (4\pi)^2} (E^*)^2  q^2 C(q) W(q) \eqno(1)$$
where $E^* = E/(1-\nu^2)$ is the effective modulus 
and where $C(q)$ is the surface roughness power spectrum, and
$$W(q) = P(q) [\gamma + (1-\gamma)P^2(q)]$$
where $P(q)$ is the relative contact area $A/A_0$ when only the roughness components with wavenumber smaller than $q$ are included in the
analysis. We have
$$P(q)={\rm erf}\left ({\sigma_0 \over 2 \surd G}\right )$$
$$G= {\pi \over 4} ( E^*)^2 \int_{q_0}^q dq \ q^3 C(q)$$
Using (1) one can show that\cite{Persson1} 
$$g({\bf x},{\bf x}') =  (E^*)^2 {1\over 4} \int d^2q  \ q^2 C(q) W(q) e^{i {\bf q} \cdot ({\bf x}-{\bf x}')}$$
Denoting $|{\bf x}-{\bf x}')| = r$ we get
$$g(r) = (E^*)^2 {\pi \over 2} \int_{q_0}^{q_1} d q  \ q^3 C(q) W(q) J_0(qr) \eqno(2)$$
As long as the relative contact area $A/A_0 \lesssim 0.1$ the contact area and $W(q)$ are proportional to $\sigma_0 /E^*$ and then (2) shows that
$g(r) =E^* \sigma_0 f(r)$ where $f(r)$ is independent of $E^*$ and $\sigma_0$. This imply that $g(r)/g(0)$ is independent of $E^*$ and $\sigma_0$ when the 
relative contact area $A/A_0 \lesssim 0.1$, which is obeyed in most rubber friction applications. We note,
however, that the integral in (2) is over all wavenumbers or magnifications $\zeta = q/q_0$, and for small magnification the surfaces appear
smooth and the relative contact area $P(q) \approx 1$. However, because of the $q^3$ factor the small wavenumber region will in most cases not dominate the integral in (2). 
{\it Since the size of the macroasperity
contact regions only depend on $g(r)/g(0)$ we conclude that the size of the macroasperity contact
regions decreases only marginally as the Young's modulus increases}. This imply
that for rubber applications, as sliding speed increases the size of the macroasperity contact regions
remains almost unchanged even though the magnitude of the viscoelastic modulus increases with sliding
speed. Furthermore, the size of the macroasperity contact regions is also nearly independent of the nominal contact pressure. 
However, these counterintuitive results are valid only if the 
nominal rubber-road contact area $A_0$ is large enough, as the theory assumes a system of infinite size.

Fig. \ref{1r.2StressCorrelation.pdf}
shows the calculated stress correlation function $g(r)$ as a function of the distance
$r=|{\bf x}-{\bf x}'|$ for rubber block squeezed against the concrete surface used in Ref. \cite{Persson1}. 
The red, green and blue curves are calculated
with the Young's modulus $E$ (in MPa) and the applied nominal contact pressure 
$\sigma_0$ (in MPa) given by $(E,\sigma_0) = (28,0.12)$ (red curve), $(14, 0.12)$ (green)
and $(14,0.24)$ (blue). The $g(r)$ for the green curve is scaled by a factor of
$2$. Eq. (2) and Fig. \ref{1r.2StressCorrelation.pdf} 
shows that when the pressure $\sigma_0$ and the Young's modulus $E$
are such that the area of real contact is proportional to $\sigma_0/E$ then
the scaled stress correlation function $g(r)/E \sigma_0$ does not depend on 
$E$ and $\sigma_0$.  

We have studied the contact between the rubber block and the concrete surface using a pressure-sensitive film (Fujifilm, Super Low-Pressure film, 
$0.5-2.5 \ {\rm MPa}$ pressure range, $\lambda = 30 \ {\rm \mu m}$ lateral resolution). Fig. \ref{FIRST.pdf}
shows the optical image of the nominal contact region of the concrete block after $5 \ {\rm s}$ of contact time. 
The nominal contact pressure $\sigma_0 = 0.12 \ {\rm MPa}$.
Fig. \ref{SECOND.pdf} shows magnified views of two contact regions from Fig. \ref{FIRST.pdf}. 
The effective (or average) diameter of the (macroasperity) contact regions is $2r_0 \sim 0.4 \ {\rm mm}$, 
which agrees beautifully with the predicted size of the macroasperity contact regions. Thus, Fig. \ref{1r.2StressCorrelation.pdf}
shows that the stress correlation function has dropped by a factor of $1/8$ at $r = 0.15 \ {\rm mm}$. We interpret the tail
of the stress correlation function for $r > 0.2 \ {\rm mm}$ to the disconnected parts of the macroasperity contact regions which
can extend away from the center of the contact region by up to $\sim 0.5 \ {\rm mm}$ (see Fig. \ref{FIRST.pdf} and \ref{SECOND.pdf}).

Fig. \ref{SandblastedAlBigWiev.pdf} and \ref{SandBlastedAl.png.pdf} shows similar contact pictures for rubber squeezed against
a sandblasted aluminum surface and Fig. \ref{SandPaperBig.pdf} and \ref{SandpaperPicMagnified.pdf} for rubber squeezed against a sandpaper P100 surface.
The result for the sandpaper is very similar to that for concrete with relative big compact contact regions surrounded by smaller disconnected contact regions.
For the sandblasted aluminum surface the contact regions are smaller and there are nearly no disconnected regions surrounding the big compact regions.
However, the density of contact areas is higher for the sandblasted surface as compared to the other two surfaces, which in part result from the
higher contact pressure used in that study. 

Fig. \ref{1logq.2logC.Concrete.Sandpaper.SandblastedAl.pdf}
shows the surface roughness power spectra of the concrete, sandpaper and sandblasted aluminum surface.
The sandpaper and the concrete both consist of mineral fragments in a binder and have similar power spectra.
The particles in sandpaper P100 have the average size $\approx 0.16 \ {\rm mm}$ while the particle sizes in the
concrete are not known to us. The power spectra was obtained from the measured line topography 
using the methods described in Ref. \cite{Nestor}.

Using these power spectra in Fig. \ref{1radius.2StressCorrelation.Concrete.sandpaper.sandblastedAl.pdf}
we show the calculated stress correlation function $g(r)$ as a function of the distance
$r=|{\bf x}-{\bf x}'|$ for an elastic block squeezed against the concrete (red line), the sandpaper (green) and the sandblasted
aluminum (blue) surfaces. We have scaled $g(r)$ for the sandpaper and sandblasted aluminum surfaces so that $g(0)$ is the same for all
three cases. Note that while the concrete and the sandpaper give similar results, with a tail extending to larger
separation $r$, this is not the case for the sandblasted surface. The size of the contact regions we observe for 
the concrete and the sandpaper, where the diameter of the compact part of the contact region is of order $0.4 \ {\rm mm}$ 
and the effective diameter when including the disconnected parts is of order $1 \ {\rm mm}$, are consistent with the 
calculated $g(r)$ if we interpret the tail as arising from the disconnected region. For the sandblasted surface
there is no such tail region which is consistent with the observed contact pictures. The diameter of the biggest contact
regions for the sandblasted surface is about $0.3 \ {\rm mm}$ which is also consistent with the calculated $g(r)$,
which describe the ensemble averages stress correlation. 


\vskip 0.3cm
{\bf 4 Summary and conclusion}

We have studied the non-uniform motion of a rubber tread block 
when the upper surface of the block is moved in a non-steady
way. We have shown that the friction coefficient experienced by the tread block 
first follows the cold-branch $\mu_{\rm cold}(v)$, where the flash temperature is negligible,
but after sliding a distance of order
the diameter $D$ of the macroasperity contact, which is needed to fully develop the flash temperature, 
it transition to the hot branch $\mu_{\rm hot} (v)$. 
Similar dynamic friction coefficients is found in tire dynamics simulations.

We have also studied the size $D$ of the macroasperity contact regions
and shown that in most cases $D$ is nearly independent of the elastic modulus and the nominal contact pressure.
This imply that for rubber friction, as the sliding speed increases the size of the contact regions are nearly unchanged
even though the magnitude of the viscoelastic modulus increases with sliding
speed. However, these counterintuitive results are valid only if the 
nominal rubber-road contact area $A_0$ is large enough, as the theory assumes a system of infinite size.

\vskip 0.2cm
{\bf Author Contributions}

Both authors have contributed equally to everything.

\vskip 0.2cm
{\bf Funding}

No funding.

\vskip 0.2cm
{\bf Data Availability} 

No datasets were generated or analyzed during the current study.

\vskip 0.2cm
{\bf Declarations}

The authors declare no competing interests.

\end{document}